\documentclass[%
 reprint,
 amsmath,amssymb,
 aps,
]{revtex4-2}

\usepackage{graphicx}% Include figure files
\usepackage{amssymb, amsmath}
\usepackage{textcomp}
\usepackage{lipsum}
\usepackage{caption}
\usepackage{subcaption}
\usepackage{float}
\usepackage{hyperref}
\usepackage{dcolumn}% Align table columns on decimal point
\usepackage{bm}% bold math

\usepackage{xcolor}
\hypersetup{
    colorlinks=true,
    allcolors=blue,
}

\begin{document}

\preprint{APS/123-QED}

\title{Study of superconductivity of very thin $\mathrm{FeSe}_{1-x}\mathrm{Te}_x$ films investigated by microwave complex conductivity measurements}% Force line breaks with \\

\author{Gaku Matsumoto}
\email{matsumoto.gaku108@gmail.com}
\author{Ryo Ogawa}
\author{Koji Higasa}
\author{Tomoki Kobayashi}
\author{Hiroki Nakagawa}
\author{Atsutaka Maeda}

\affiliation{%
 Department of Basic Science, University of Tokyo
}%

\begin{abstract}
Complex conductivity measurements spanning the entire temperature range, including the vicinity of $T_c$, were conducted on systematically varied FeSe$_{1-x}$Te$_x$ ($x$ = 0 - 0.5) very thin films. By applying a novel cavity measurement technique employing microwave electric fields parallel to FeSe$_{1-x}$Te$_x$ films, we observed distinct temperature-dependent alterations in superfluid fraction and quasiparticle scattering rate at the nematic boundary. These changes in the nematic boundary suggests variations in the superconducting gap structure between samples in the nematic and non-nematic phase. Moreover, fluctuation is visible up to 1.2 $T_c$ irrespective of nematic order, consistent with large superconducting fluctuations in iron chalcogenide superconductors reported previously in [H. Takahashi \textit{et al}, Phys. Rev. B 99, 060503(R) (2019)] and [F. Nabeshima \textit{et al}, Phys. Rev. B 97, 024504(R) (2018)].
\end{abstract}

\maketitle

\section{Introduction}
Iron-based superconductors have been the subject of intensive research from both fundamental and applied physics perspectives due to their high superconducting transition temperature under ambient pressure, second only to copper oxide superconductors\cite{Liu,Kri}. Among iron-based superconductors, an iron chalcogenide superconductor $Fe\it{Ch}$ ($\it{Ch}$=S, Se, Te) are known for their simple crystal structure\cite{Hsu}. The iron chalcogenide superconductor, $\mathrm{FeSe}$, has garnered significant attention due to its distinctive properties. These include the potential for high temperature superconductor, its multiple-bands nature, the absence of magnetic order under ambient pressure, its small Fermi surface with a superconducting gap of similar magnitude\cite{Kri,shiba}. Various methods, such as intercalation\cite{intercalation2, intercalation3}, carrier doping using electron double layer transistor\cite{edlt1,edlt2,edlt3,edlt4}, and the synthesis of monolayer films\cite{mono1,mono4}, have been employed to significantly raise the superconducting transition temperature, $T_{\mathrm{c}}$, in $\mathrm{FeSe}$ from 8 K to more than 65 K\cite{mono4}. The nematic phase of $\mathrm{FeSe}$\cite{Fer,nem,Rosler}, which lacks magnetic order, provides an excellent opportunity to explore not only the origin of nematicity but also its relationship with superconductivity. Besides, $\mathrm{FeSe}$ is known for its small Fermi surface ($\mathrm{\epsilon_F}$ $<$ 10 meV)\cite{Lubashevsky, Okazaki, kasahara,fsss}. Since the changes in Fermi surface impact the superconducting, nematic, and magnetic phases, a variety of techniques have been explored to investigate the electronic phase of $\mathrm{FeSe}$ and its exotic superconductivity. Its Fermi surface can be easily adjusted through chemical substitution\cite{Hsu,chemicalpressureiso1,chemicalpressureiso2,chemicalpressureiso3}, hydrostatic pressure\cite{hydrostatic}, and lattice strain in the plane\cite{lattice}.  Among these methods mentioned to manipulate the electronic state, chemical isovalent substitution stands out as advantageous, which has been commonly practiced. When Tellurium (Te) is substituted into $\mathrm{FeSe}$, it induces lattice expansion, effectively creating negative chemical pressure\cite{fsun, Te}.

Despite the issue of phase separation being prone to occur in bulk samples  of $\mathrm{FeSe}_{1-x}\mathrm{Te}_x$ with a composition range of $\it{x}$ = 0.1 
- 0.4\cite{chemicalpressureiso1, difficulty3, difficulty4}, the question of how superconductivity changes with increasing Te content remains a matter of significant importance. Before systematically producing bulk $\mathrm{FeSe}_{1-x}\mathrm{Te}_x$\cite{terao}, we successfully grew single-crystalline thin films of $\mathrm{FeSe}_{1-x}\mathrm{Te}_x$ spanning the entire composition range ($\it{x}$ = 0 - 0.9) using a pulsed laser deposition technique\cite{chemicalpressureiso3,phasesuccess2}. Intriguingly, despite the decrease in the nematic transition temperature, $T_{\mathrm{n}}$, observed in $\mathrm{FeSe}_{1-x}\mathrm{Te}_x$ films after Te substitution, there was a substantial increase in $T_{\mathrm{c}}$ following the disappearance of nematic order\cite{chemicalpressureiso3,phasesuccess2}. This significant enhancement of $T_{\mathrm{c}}$ stands in contrast to the behavior observed in $\mathrm{FeSe}_{1-x}\mathrm{Te}_x$ bulk\cite{Ishida,terao}, $\mathrm{FeSe}_{1-x}\mathrm{S}_x$ bulk\cite{FeSeSbulk, Ishida} and $\mathrm{FeSe}_{1-x}\mathrm{S}_x$ film\cite{nabe}, suggesting that the influence of nematicity on $T_{\mathrm{c}}$ is complicated within these materials. Furthermore, it was found that in the low temperature region, there are changes in the temperature dependencies of penetration depth and quasiparticle scattering rate at the nematic boundary obtained from complex conductivity, as reported in \cite{kurokawa}. However, the complex conductivity of thin film remains unexplored in the vicinity of $T_{\mathrm{c}}$ as of yet.\par

In this paper, we report a novel method for measuring the complex conductivity, $\mathrm{\sigma}$, in the vicinity of the critical temperature in very thin films, a technique that has not been established thus far. By 'very thin,' we mean that the film thickness, $d$, is significantly thinner than the penetration depth, $\lambda$, i.e., $d \ll \lambda$. We present the establishment of this method, utilizing a microwave electric field applied parallel to the film surface, to measure complex conductivity near $T_{\mathrm{c}}$, and we applied this technique to $\mathrm{FeSe}_{1-x}\mathrm{Te}_x$ ($x$ = 0 - 0.5) films to investigate their superconducting characteristics. To evaluate both the real and imaginary parts of $\mathrm{\sigma}$ in the higher temperature region including the vicinity of $T_{\mathrm{c}}$, we integrated microwave cavity perturbation techniques with the mutual inductance method. The obtained complex conductivity exhibited characteristic behavior typical of high-temperature superconductors. The quasiparticle scattering rate was calculated from the real part of $\mathrm{\sigma}$ assuming the Drude model, and was found to decrease drastically from the vicinity of $T_{\mathrm{c}}$, as observed in bulk $\mathrm{FeSe}$\cite{oka} and $\mathrm{FeSe}_{0.4}\mathrm{Te}_{0.6}$\cite{taka}. Furthermore, the temperature dependent behavior of quasiparticle scattering rate and superfluid density in a series of $\mathrm{FeSe}_{1-x}\mathrm{Te}_x$ films with varying $\it{x}$ values exhibited distinct differences between the nematic and non-nematic phases even around $T_{\mathrm{c}}$. This suggest that the gap structure is different  between two groups, which is  consistent with measurements using ARPES, which indicate that the electronic state of the normal conducting state changes before and after the nematic transition\cite{ARPES}. The superconducting fluctuation remains observable up to 1.2 $T_{\mathrm{c}}$, irrespective of the presence or absence of nematic order, consistent with previous reports employing microwave spectroscopy\cite{sflu1} or precise torque magnetometry\cite{sflu2}.\par

\section{EXPERIMENTS}

\subsection{Sample}
All $\mathrm{FeSe}_{1-x}\mathrm{Te}_x$ films were deposited onto CaF$_2$ substrates using a pulsed laser deposition method with a KrF laser. The details of film growth were described elsewhere\cite{PLD1,PLD2}. The thickness of the thin films used in this study ranged from 50 nm to 80 nm, as measured using a stylus profiler or interferometry. For measurements using the mutual inductance method, samples were fabricated in a circular shape with a diameter of 6.7 mm, and subsequently cut to dimensions of 0.3 mm in length, 0.1 mm in width, and 0.5 mm in height including substrate for cavity resonator measurements. The dc resistivity was measured using a four-probe method equipped with a physical property measurement system (Quantum Design, PPMS).

\subsection{Measurements and Analysis}
The complex conductivity of $\mathrm{FeSe}_{1-x}\mathrm{Te}_x$ ($\it{x}$ = 0 - 0.5) films was measured using the cavity perturbation technique. For very thin films $(d\ll\lambda)$, conventional cavity perturbation measurements in microwave magnetic fields\cite{Maereview} are restricted to low-temperature ranges, spanning from the lowest temperature to at most approximately 0.7 $T_{\mathrm{c}}$\cite{kurokawa}. This limitation arises because the electromagnetic field configuration undergoes significant changes depending on the temperature region, either $T\geq T_{\mathrm{c}}$ and $T\ll T_{\mathrm{c}}$, as elucidated by Barannik \textit{et al}\cite{Barannik}.  Consequently, we employed a cavity perturbation technique utilizing microwave electric fields parallel to the film surface to measure complex conductivity in the vicinity of $T_{\mathrm{c}}$. 
\begin{figure}[htbp]
    \begin{tabular}{cc}
      %---- 最初の図 ---------------------------
      \begin{minipage}[t]{0.343069\hsize}
        \centering
        \includegraphics[keepaspectratio, width=\linewidth]{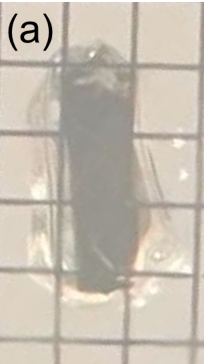}
        \label{fig:flake}
      \end{minipage} &
      %---- 2番目の図 --------------------------
      \begin{minipage}[t]{0.62376\hsize}
        \centering
        \includegraphics[keepaspectratio, width=\linewidth]{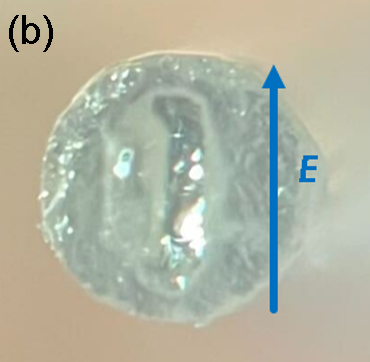}
        \label{fig:ttt}
      \end{minipage}
      %---- 図はここまで ----------------------
    \end{tabular}
    \captionsetup{justification=raggedright, singlelinecheck=false}
    \caption{The sample of $\mathrm{FeSe}_{0.6}\mathrm{Te}_{0.4}$, cut to approximately 0.3 × 0.1 × 0.5 mm³ in size, was prepared for measurement using the cavity perturbation method. (a) This image was captured using an optical microscope, with the sample placed on graph paper with a grid spacing of 0.1 mm per square. (b) This is a photograph of a sample placed on a sapphire rod. The diameter of the top surface of the sapphire rod is 0.5 mm. The blue arrow indicates the direction of the applied electric field.}
    \label{fig:dd}
  \end{figure}
  
A flake of $\mathrm{FeSe}_{1-x}\mathrm{Te}_x$ film with substrate was cut into approximately 0.3 × 0.1 × 0.5 mm³ in size as depicted in Fig. \ref{fig:dd}\hyperref[fig:dd]{(a)}. This flake was affixed onto a sapphire rod positioned at the location with magnetic field nodes and a notably intense electric field within the resonator as depicted in Fig. \ref{fig:dd}\hyperref[fig:dd]{(b)}. The resonator operated in the $\mathrm{TE}_{011}$ mode at 44 GHz, configured such that the electric field of the $\mathrm{TE}_{011}$ mode was aligned parallel to the film surface, thereby allowing surface current to flow directly within the ab plane. In the cavity pertubation  method, changes in the resonant characteristics, the resonant frequency, $f$, and the quality factor of the cavity, $Q$, are measured. The complex frequency shift is expressed as follows:  
\begin{equation}
\begin{aligned}
\frac{\Delta \omega}{\omega} &= \frac{\Delta f}{f} - i\Delta\frac{1}{2 Q}\\
&\equiv \frac{f_\mathrm{sample}-f_\mathrm{blank}}{f_\mathrm{sample}}-i(\frac{1}{2Q_\mathrm{sample}}- \frac{1}{2Q_\mathrm{blank}}),
\end{aligned}
\end{equation}
where $Q_\mathrm{sample}$ and $f_\mathrm{sample}$ represent the quality factor and the resonant frequency of the cavity containing the sample, respectively, while $Q_\mathrm{blank}$ and $f_\mathrm{blank}$ denote the quality factor and the resonant frequency of the cavity in the absence of the sample, respectively.
The complex frequency shift is correlated with the complex conductivity, $\sigma$, expressed in the following equation derived by Peligrad $\it{et}$ $\it{al}$\cite{Peligrad}.
\begin{equation}
 \begin{aligned}
& \frac{\Delta \omega}{\omega}=\frac{\Gamma}{N}\left[\left\{1+\left(\frac{k^2}{k_0^2}-1\right) N\right\} \frac{\tanh (\mathrm{i} k d / 2)}{\mathrm{i} k d / 2}\right]^{-1},\\
& \text { where } k=k_0 \sqrt{\mu_{\mathrm{r}}\left(1-\mathrm{i} \frac{\sigma}{\epsilon_0 \omega}\right)}, \quad k_0=\omega \sqrt{\epsilon_0 \mu_0}.
\end{aligned}
  \label{eq:cpm}
\end{equation}
Here, $N$ represents the depolarization factor, $\it{\Gamma}$ is the filling factor, $\it{\epsilon_0}$ denotes the permittivity in vacuum, $\mu_0$ stands for the permeability in vacuum, and $\mu_r$ signifies the permeability in a substance and $\omega$ is angular frequency. The magnetic penetration length at each temperature, $\mathrm{\lambda}$, was measured to determine the unknown factors, $\it{\Gamma}$ and $\it{N}$, in the following manner.

To measure the absolute value of $\mathrm{\lambda}$$(\it{T})$ of $\mathrm{FeSe}_{1-x}\mathrm{Te}_x$ films, we employed the two-coil mutual inductance technique\cite{mi1,mi2}, which measures the mutual inductance, denoted as $M$, between pickup coil and drive coil. The measurements conducted in this study are based on the following principles. First, consider a situation where each coil is a single winding, with a coil-to-coil distance of $D$, and the film in between has infinite extent and thickness $\it{d}$. We assumed a drive coil with an inner radius of $R_\mathrm{d}$ and a pickup coil with an inner radius of $R_\mathrm{p}$. In this situation, Clem and Coffey have shown that the mutual inductance is expressed as follows\cite{cc}.
\begin{equation}
M_{j k}\left(\lambda, \sigma_1\right) = \mu_0 \pi R_{\mathrm{d}} R_{\mathrm{p}} \int_0^{\infty} \frac{e^{-q D} J_1\left(q R_{\mathrm{d}}\right) J_1\left(q R_{\mathrm{p}}\right)}{\cosh\left(S d\right) + \frac{S^2+q^2}{2 S q \sinh\left(S d\right)}} \, dq.
\label{eq:mim}
\end{equation}
Here, $q$ is the wave-number, $J_1$ represents the first-order Bessel function, and we define $S = q^2 + \lambda^{-2} + \mathrm{i} \mu_0 \omega \sigma_1$ where $\sigma=\sigma_1-$i$\sigma_2$.
The mutual inductance of the multi-layered coil is represented by the sum of all combinations of the Eq. (\ref{eq:mim}) as follows.
\begin{equation}
M\left(\lambda, \sigma_1\right)=\sum_{j, k} M_{j k}\left(\lambda, \sigma_1\right).
\label{eq:oppp}
\end{equation}
Moreover, while the above situation considers a superconducting thin film extending infinitely, thin films used in actual measurements have finite sizes. As a result, magnetic fields leaking from the sample edges induce excess magnetic coupling, leading to the generation of leakage mutual inductance $M^{\text{leak}}$. The deviation for this situation has been analyzed by Turneture \textit{et al.}\cite{Turneature}, and it has been shown that the following relationship holds.
\begin{equation}
    M^{\text{meas}}\simeq M\left(\lambda, \sigma_1\right)+M^{\text{leak}}.
    \label{eq:leak}
\end{equation}
$M^{\text{meas}}$ represents the mutual inductance actually measured. On the other hand, $M^{\text{leak}}$ depends solely on the sample size on the plane and can be evaluated by measuring the mutual inductance of a superconducting film with the same radius as the measurement sample but sufficiently thicker, i.e., completely shielding the electromagnetic field\cite{Classen}. To realize the situation where the above Eq. (\ref{eq:leak}) is applicable, we utilized circular sample with a radius of 6.7 mm in this series of experiments. To estimate the magnitude of $M^{\text{leak}}$, measurement results from a 1 $\mathrm{{\mu}m}$ thick NbN film were performed.

Since the measured mutual inductance is represented by the sum of $M_{jk}$ over all possible combinations $(j, k)$ as described in Eq. (\ref{eq:oppp}), evaluating the magnetic penetration depth, $\lambda$, from the measured mutual inductance, $M$, entails solving the inverse problem for Eq. (\ref{eq:oppp}), treating it as a function that given $M$, returns $\lambda$ and $\sigma_1$. The procedure for solving this inverse problem, similar to the method by He \textit{et al}\cite{Bozobitch}, is as follows:

\begin{enumerate}
   \item Create a table of M($\lambda$, $\sigma_1$) using numerical computations.
   \item Within the table, identify the value M(1) that is the closest to the experimental result M(exp).
   \item Create a smaller table centered around M(1) and determine the optimal value M(2) within this table.
   \item Repeat the same procedure, iterating until $\lambda$ and $\sigma_1$ corresponding to M(10) are obtained as the computed results.
\end{enumerate}

In this manner, the magnetic penetration depth $\lambda$(\textit{T}) can be determined across a wide temperature range from the values of mutual inductance.

\section{RESULTS AND DISCUSSION}\par
\begin{figure}[htbp]
    \begin{subfigure}[t]{0.47\textwidth}
        \centering
        \includegraphics[keepaspectratio, width=\linewidth]{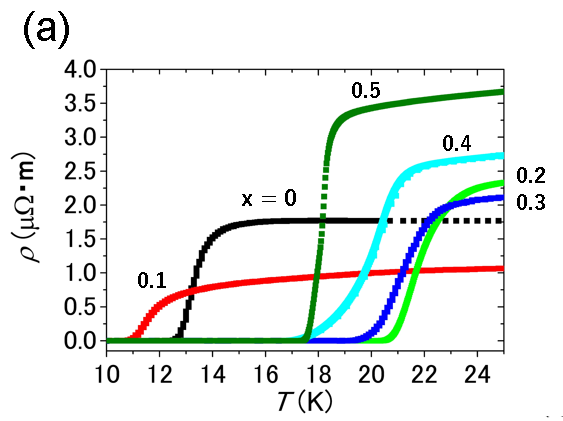}
        \label{fig:flake}
    \end{subfigure}
    \begin{subfigure}[t]{0.475\textwidth}
        \centering
        \includegraphics[keepaspectratio, width=\linewidth]{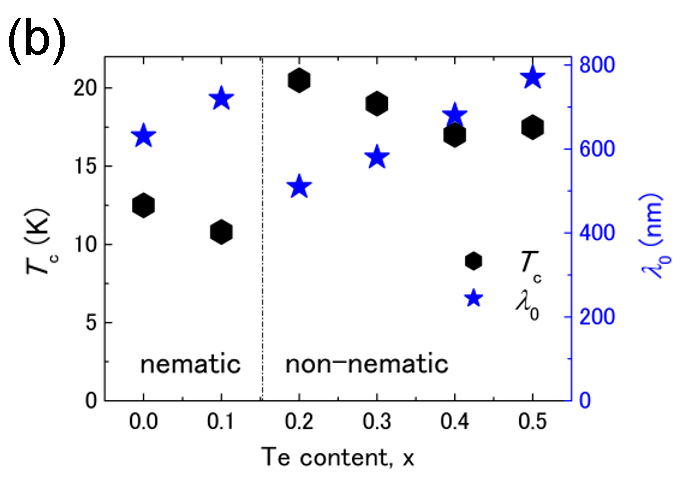}
        \label{fig:ttl}
    \end{subfigure}
    \captionsetup{justification=raggedright, singlelinecheck=false}
    \caption{(a) Temperature dependence of dc resistivity of the $\mathrm{FeSe}_{1-x}\mathrm{Te}_x$ ($\it{x}$ = 0$–$0.5) films. (b) Te content $\it{versus}$ $T_{\mathrm{c,zero}}$ and $\mathrm{\lambda}_{{0}}$ utilized in subsequent complex conductivity measurement. The hexagons represent $T_{\mathrm{c,zero}}$ obtained from dc resistivity measurements, while the stars represent the magnetic penetration depth at 0 K.}
    \label{fig:dd2}
\end{figure}
Fig. \ref{fig:dd2}\hyperref[fig:dd2]{(a)} shows the temperature dependence of dc resistivity in $\mathrm{FeSe}_{1-x}\mathrm{Te}_x$ ($\it{x}$ = 0$–$0.5) films. When systematically varying $\it{x}$, $T_{\mathrm{c}}$ significantly increases in the composition where the nematic phase disappears. Conversely, as $\it{x}$ is increased, $T_{\mathrm{c}}$ gradually decreases in the non-nematic phase. This trends align with previous reports\cite{phasesuccess2}.

We utilized the temperature dependent values of  the mutual inductance $M$ obtained through the mutual inductance method to calculate $\mathrm{\lambda}(T)$ using Eq. (\ref{eq:mim}) and Eq. (\ref{eq:leak}). The $\mathrm{\lambda}(T)$ values obtained were extrapolated to 0 K, assuming that $\mathrm{\lambda}(T)$ follows the form $\mathrm{\lambda}_{{0}}+ A(T/T_{\mathrm{c}})^n$, where $\mathrm{\lambda}_{{0}}$ represents the penetration depth at 0 K, and $A$ and $n$ are constants. To accomplish this, we conducted curve fitting within the temperature range from 5 K to the lowest temperature ($\sim$ 2 K) applicable for samples that could be analyzed using the mutual inductance method. Fig. \ref{fig:dd2}\hyperref[fig:dd2]{(b)} illustrates the relationship between Te content and $T_{\mathrm{c,zero}}$, as well as $\mathrm{\lambda}_{{0}}$, which are employed in subsequent complex conductivity measurements.

\begin{figure}[htbp]
    \begin{subfigure}[t]{0.45\textwidth}
        \centering
        \includegraphics[keepaspectratio, width=\linewidth]{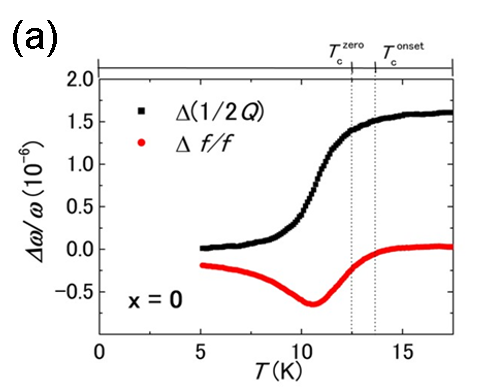}
        \label{fig:flake}
    \end{subfigure}
    \begin{subfigure}[t]{0.482\textwidth}
        \centering
        \includegraphics[keepaspectratio, width=\linewidth]{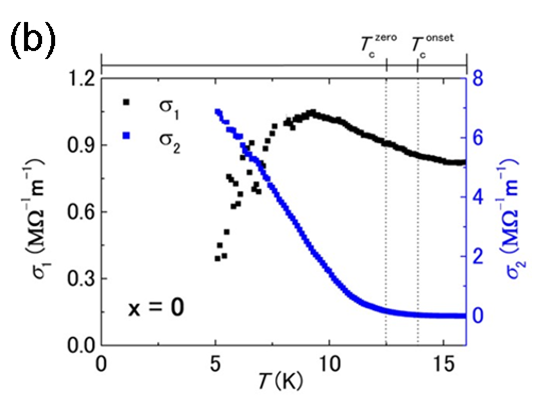}
        \label{fig:ttl}
    \end{subfigure}
    \captionsetup{justification=raggedright, singlelinecheck=false}
    \caption{(a) Temperature dependence of the complex frequency shift of the $\mathrm{FeSe}$ film. (b) Temperature dependence of the complex conductivity of the $\mathrm{FeSe}$ film obtained from the data in Fig. \ref{fig:dd4}\hyperref[fig:dd4]{(a)}.}
    \label{fig:dd4}
\end{figure}

Subsequently, we assessed the dynamics of quasiparticles and superfluidity utilizing the cavity perturbation technique under microwave electric fields. In Fig. \ref{fig:dd4}\hyperref[fig:dd4]{(a)}, the temperature dependence of the shift of $Q^{-1}$ and $f$ of the $\mathrm{FeSe}$ film was shown as a representative. The obtained result shows qualitative agreement with the behavior of the complex frequency shift obtained under the assumption of a two-fluid model, as calculated by Peligrad \textit{et al.}\cite{Peligrad}.  Here, we confirmed that the effect of the $\mathrm{CaF_2}$ substrate was negligible by the measurement of the substrate alone.

From the measurement of the $Q^{-1}$(${T}$) and $f$(${T}$) in each films placed in the cavity resonator, the complex conductivity were calculated using Eq. (\ref{eq:cpm}). Fig. \ref{fig:dd4}\hyperref[fig:dd4]{(b)} shows $\mathrm{\sigma}(T)$ of the $\mathrm{FeSe}$ film. The real part, $\mathrm{\sigma}_1$, exhibits a broad peak from near $T_{\mathrm{c}}$ down to low temperatures. The $\mathrm{\sigma}_1$ behavior and its magnitude are in good agreement with results obtained in bulk crystals\cite{taka}. On the other hand, the imaginary part, $\mathrm{\sigma}_2$, starts to acquire values greater than zero near $T_{\mathrm{c}}$ and monotonically increases as the temperature decreases.

\begin{figure}[hbtp]
 \centering
 \includegraphics[keepaspectratio, scale=0.2641]
      {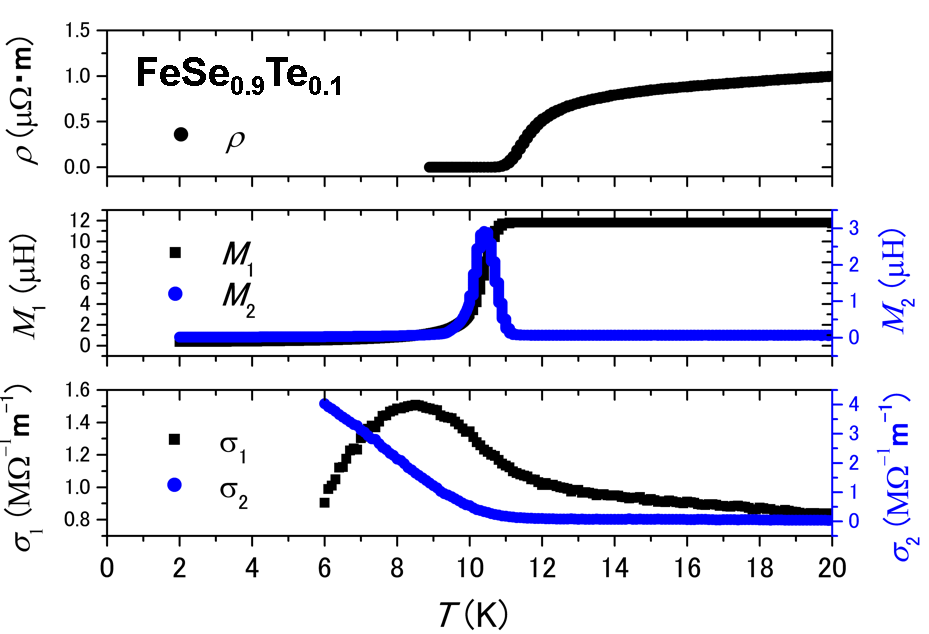}
 \captionsetup{justification=raggedright, singlelinecheck=false}
 \caption{The dc resistivity at the top, ac magnetization at 30 kHz in the middle, and the complex conductivity measured at 44 GHz at the bottom is shown here. These were all measured for the same $\mathrm{FeSe}_{0.9}\mathrm{Te}_{0.1}$ film.}
 \label{fig:cfs}
\end{figure}

The relationship between microwave complex conductivity and fundamental quantities including dc resistivity and diamagnetic susceptibility of the $\mathrm{FeSe}_{0.9}\mathrm{Te}_{0.1}$ film is illustrated in the graphical representation provided in Fig. \ref{fig:cfs}, derived from measurements conducted via the mutual inductance method described previously. Within Fig. \ref{fig:cfs}, arranged from top to bottom, are representations of dc resistivity, ac magnetization at 30 kHz, and complex conductivity obtained at 44 GHz. According to this correlation, the response of ac magnetization is observed at the temperature where dc resistivity approaches zero, signaling the onset of an increase in $\mathrm{\sigma}_2$. Notably, a significant increase in temperature dependence of $\mathrm{\sigma}_1$ is discerned as the dc resistivity diminishes during the superconducting transition. We will discuss the details later.

\begin{figure}[htbp]
    \begin{subfigure}[t]{0.48\textwidth}
        \centering
        \includegraphics[keepaspectratio, width=\linewidth]{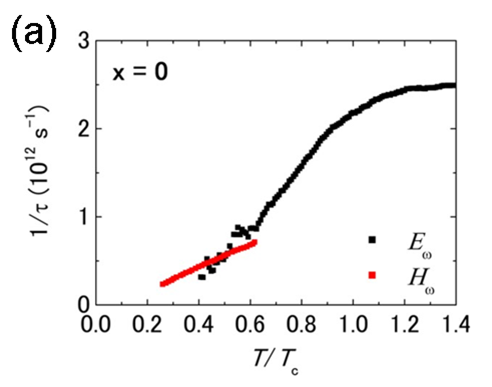}
        \label{fig:flake}
    \end{subfigure}
    \begin{subfigure}[t]{0.48\textwidth}
        \centering
        \includegraphics[keepaspectratio, width=\linewidth]{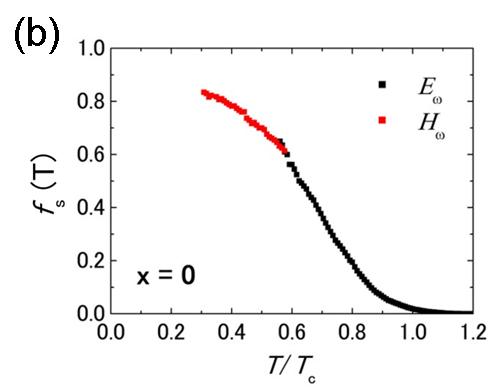}
        \label{fig:ttl}
    \end{subfigure}
    \captionsetup{justification=raggedright, singlelinecheck=false}
    \caption{(a) Temperature dependence of the quasiparticle scattering rate of the $\mathrm{FeSe}$ films. The black dots here represent the data measured using the microwave electric field, while the red dots in the figure represent the low-temperature data measured using the microwave magnetic field\cite{kurokawa}. (b)  Temperature dependence of the superfluid density of the $\mathrm{FeSe}$ film. Also, the black dots in the figure represent the data measured using the microwave electric field, while the red dots in the figure represent the low-temperature data measured using the microwave magnetic field\cite{kurokawa}.}
    \label{fig:opp9}
\end{figure}

Assuming the two-fluid model and a Drude-like single-carrier normal fluid, we can express the quasiparticle scattering time, $\tau$, as follows:
\begin{equation}
\omega \tau=\frac{\tilde{\sigma_1}}{1-\tilde{\sigma_2}}
  \label{eq:drude}
\end{equation}
where $\tilde{\sigma}\equiv\tilde{\sigma}_1-{i} \tilde{\sigma}_2=\mu_0 \omega \lambda_0^2\left(\sigma_1-i \sigma_2\right)$, representing the dimensionless conductivity. While it is well-known that multiple carriers are involved in superconductivity, we deliberately proceeded the analysis considering a single band, assuming that the temperature dependence of $\tau$ in both hole and electron pockets could be represented using a single $\tau$ as in previous studies\cite{kurokawa}. As the temperature dependence of $\tau$ remained unaffected considerably by the values of residual surface resistance, we used $\tau$ rather than $\mathrm{\sigma}_1$ to examine the intrinsic characteristics of $\mathrm{FeSe}_{1-x}\mathrm{Te}_x$ films. First, we discuss the quasiparticle scattering rate for a specific Te content, in this case, for $x$ = 0. Fig. \ref{fig:opp9}\hyperref[fig:opp9]{(a)} illustrates the temperature dependence of quasiparticle scattering rates in $\mathrm{FeSe}$ films. The black plots in the Fig. \ref{fig:opp9}\hyperref[fig:opp9]{(a)} represent the data measured using the microwave electric field. It is evident that quasiparticle scattering rates decrease significantly from the vicinity of $T_{\mathrm{c}}$. This reduction corresponds to the suppression of inelastic scattering processes, such as carrier-carrier scattering, not necessarily limited to electron-electron scattering, accompanying the opening of the superconducting gap. This trend represents a common behavior observed when inelastic scattering predominates at the superconducting transition temperature. The red plots in Fig. \ref{fig:opp9}\hyperref[fig:opp9]{(a)} represent the low-temperature data measured using the microwave magnetic field\cite{kurokawa}. Both black and red plots agree well, which means that a temperature dependence spanning the entire temperature range has been obtained successfully. 

Next, we discuss the superfluid density. When assuming that at 0 K, all electrons condense into a superconducting state as Eq. (\ref{eq:ass}), the temperature dependence of the superfluid density is described as Eq. (\ref{eq:superfluid}).
\begin{equation}
n_{\mathrm{s}}(T)+n_{\mathrm{n}}(T)=n_{\mathrm{s}}({0})
  \label{eq:ass}
\end{equation}

\begin{equation}
f_s(T)\equiv \frac{n_{\mathrm{s}}(T)}{n_{\mathrm{s}}({0})}=\tilde{\sigma_2}-\frac{{\tilde{\sigma_1}}^2}{1-\tilde{\sigma_2}},
  \label{eq:superfluid}
\end{equation}
where $f_s(T)$ represents superfluid fraction.
Fig. \ref{fig:opp9}\hyperref[fig:opp9]{(b)} shows the temperature-dependent superfluid fraction of the $\mathrm{FeSe}$ film. Again, the black plots in Fig. \ref{fig:opp9}\hyperref[fig:opp9]{(b)} represent the data measured using the microwave electric field, and the red plots in Fig. \ref{fig:opp9}\hyperref[fig:opp9]{(b)} represent the low-temperature data measured using the microwave magnetic field\cite{kurokawa}. In this manner, the results of the superfluid not only validate the accuracy of the measurements but also enable data acquisition across a wide temperature range. $f_s(T)$ exhibits a monotonous increase from near $T_{\mathrm{c}}$ to the lowest temperature and tends to saturate as it approaches 0 K. Here again, the black plots and the red plots connect smoothly.

\begin{figure}[htbp]
    \begin{subfigure}[t]{0.48\textwidth}
        \centering
        \includegraphics[keepaspectratio, width=\linewidth]{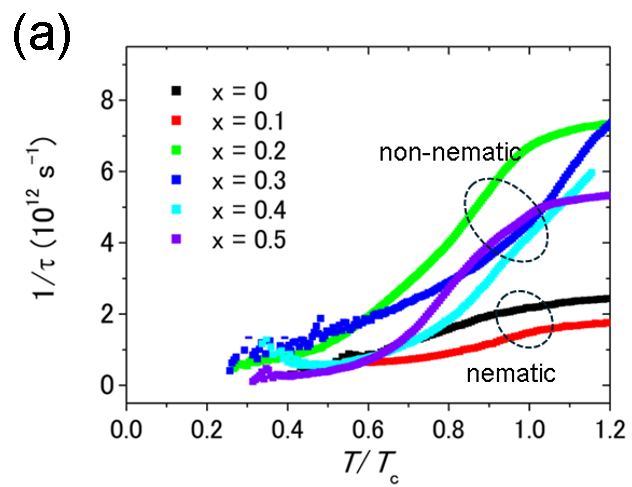}
        \label{fig:flake}
    \end{subfigure}
    \begin{subfigure}[t]{0.48\textwidth}
        \centering
        \includegraphics[keepaspectratio, width=\linewidth]{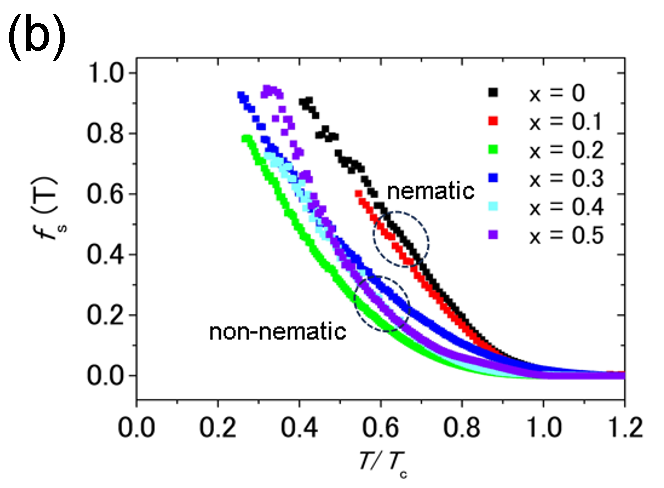}
        \label{fig:ttl}
    \end{subfigure}
    \captionsetup{justification=raggedright, singlelinecheck=false}
    \caption{(a) Quasiparticle scattering rate for $\mathrm{FeSe}_{1-x}\mathrm{Te}_x$ with systematically varying Te content (b) Temperature-dependent superfluid density for $\mathrm{FeSe}_{1-x}\mathrm{Te}_x$ samples with varying $\it{x}$ values}
    \label{fig:crazy}
\end{figure}

Next, we discuss the dependence of the quasiparticle scattering rate and superfluid fraction on the Te content. Fig. \ref{fig:crazy}\hyperref[fig:crazy]{(a)} depicts the quasiparticle scattering rate for $\mathrm{FeSe}_{1-x}\mathrm{Te}_x$ with systematically varying Te content. The slope of rate differs between samples in nematic regions and non-nematic regions. Samples in the nematic region show small slopes, while samples in the non-nematic region exhibit larger slopes, which indicates that scattering mechanisms differ between the two groups.

Fig. \ref{fig:crazy}\hyperref[fig:crazy]{(b)} displays the temperature dependent superfluid fraction for a range of $\mathrm{FeSe}_{1-x}\mathrm{Te}_x$ samples with varying $\it{x}$ values. Again, a clear distinction emerges when comparing samples within the nematic ($\it{x}$ = 0$–$0.1) and non-nematic ($\it{x}$ = 0.2$–$0.5) regions; the slope of $f_{\mathrm{s}}(T)$ appears to be different among the two groups. In the nematic phase, it shows a linear temperature dependence of the superfluid density from 0.9 $T_{\mathrm{c}}$ to the lowest temperature, whereas, in the non-nematic phase, it exhibits a temperature dependence that can be expressed as a sum of non-linear power laws in superconducting regions. Moreover, it can be observed that in the nematic region, the group exhibits a higher superfluid density than in the non-nematic region in this temperature range. In other words, the superfluid exhibits a more rapid growth in the nematic region compared to the non-nematic region near $T_{\mathrm{c}}$, suggesting a discrepancy in the superconducting gaps between the two groups.

In this series of experiments, in both the quasiparticle scattering rate and superfluid fraction, distinct differences were observed between the nematic and non-nematic groups, suggesting different gap structure between the two groups. This observation aligns well with the results obtained from complex conductivity measurements utilizing magnetic field components\cite{kurokawa}. Moreover, this is consistent with the results from ARPES\cite{ARPES}, which have shown that substitution of Te significantly alters the electronic states near the Fermi level. According to this report, in the nematic state, only the $d_{yz}$ and $d_{xz}$ orbitals are near the Fermi level, whereas substitution with Te causes the energy of the $d_{xy}$ orbital to increase, leading to hybridization with electronic states forming the Fermi surface. Consequently, the density of states near the Fermi level undergoes significant changes. Furthermore, not only ARPES measurements\cite{ARPES} but also carrier dynamics\cite{carrierdynamics1,carrierdynamics2}, optical spectroscopy\cite{optical1,optical2}, and DFT calculations\cite{kurokawa} suggest changes in electronic states at the nematic boundary. Considering that in thin films, a pure nematic transition, which is a nematic transition that does not involve lattice transitions, occurs\cite{pure}, it can be concluded this difference in this gap structure arises from the pure response of the bare electronic system. Therefore, the sharp change in $T_{\mathrm{c}}$ at the nematic boundary in $\mathrm{FeSe}_{1-x}\mathrm{Te}_x$ thin films, which is not seen in bulk samples\cite{Ishida}, is solely attributed to this change in electronic structure.

\begin{figure}[htbp]
    \includegraphics[width=1.0\linewidth]{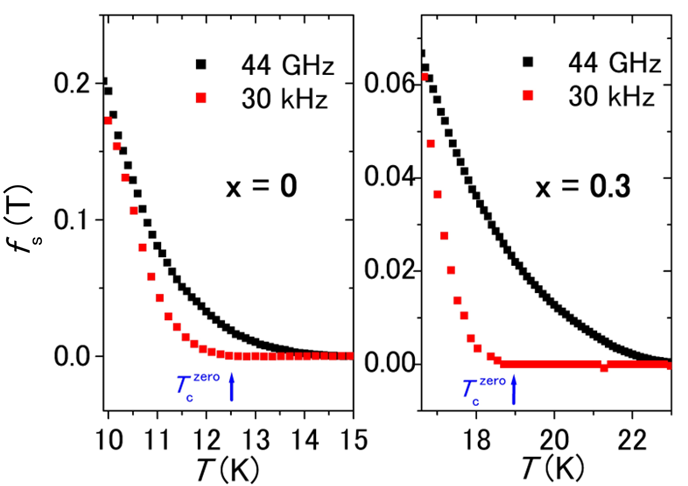}
    \captionsetup{justification=raggedright, singlelinecheck=false}
    \caption{Superfluid fraction of FeSe (on the left panel) and FeSe$_{0.7}$Te$_{0.3}$ (on the right panel) at 44 GHz (black plots) calculated from complex conductivity and 30 kHz (red plots) obtained from the mutual inductance method.}
    \label{fig:f95}
\end{figure}

Finally, we discuss the superconducting fluctuations obtained from complex conductivity measurements near $T_{\mathrm{c}}$. Fig. \ref{fig:f95} compares the superfluid fraction at two different frequencies. The one at 44 GHz is obtained from the complex conductivity, while the one at 30 kHz is obtained using the aforementioned mutual inductance method. It can be seen that the superfluid fraction measured at 30 kHz starts to rise from nearly $T_{\mathrm{c}}^{zero}$, where dc resistivity becomes zero. On the other hand, the superfluid appears at higher temperatures at 44 GHz. As expected, a more pronounced superfluid response is observed at higher frequencies, extending to higher temperatures. It can be stated that superconductivity fluctuations are observable up to about 1.2 times $T_{\mathrm{c}}^{zero}$ in both nematic and non-nematic samples. This is consistent with our prior investigations on FeSe$_{0.5}$Te$_{0.5}$ measured by using microwave spectroscopy ($\sim$ 1.12 $T_\mathrm{c}$)\cite{sflu1} or FeSe measured on precise magnetic torque measurement(at most 1.2 $T_\mathrm{c}$ in the low magnetic field region)\cite{sflu2}, providing further proof of large superconducting fluctuations in iron chalcogenides.

\section{CONCLUSION}
In conclusion, we developed a cavity measurement technique using microwave electric fields to measure complex conductivity of very thin films in the vicinity of $T_{\mathrm{c}}$ and employed this newly developed cavity pertubation technique using microwave electric field to epitaxial FeSe$_{1-x}$Te$_x$ films ($\it{x}$ = 0 - 0.5). It was confirmed that the temperature dependence of superfluid density and quasiparticle scattering time changes across the nematic boundary. Similar to magnetic field measurement\cite{kurokawa}, this suggests that the superconducting gap structure differs between samples in the nematic and non-nematic phases. This corresponds to the change in electronic states at the nematic boundary, solely accounting for the sharp increase in $T_{\mathrm{c}}$. Furthermore, fluctuation is visible up to 1.2 $T_c$ irrespective with or without nematic order, consistent with large superconducting fluctuations in iron chalcogenide superconductors\cite{sflu1,sflu2}.

\end{document}